\documentclass[%
 aip,jap,
 amsmath,amssymb,
 reprint,%
]{revtex4-1}

\usepackage{graphicx}
\usepackage{dcolumn}
\usepackage{bm}
\usepackage[utf8]{inputenc}
\usepackage[T1]{fontenc}
\usepackage{mathptmx}

\begin{document}
\preprint{AIP/123-QED}

\title{Crystal Hall and crystal magneto-optical effect  
in thin films of SrRuO$_3$
}

\author{Kartik Samanta}
\affiliation{Peter Grunberg Institut and Institute for Advanced Simulation, 
Forschungszentrum J\"{u}lich and JARA, 52425 J\"{u}lich, Germany}
\email{k.samanta@fz-juelich.de}
\author{Marjana Ležaić}
\affiliation{Peter Grunberg Institut and Institute for Advanced Simulation, 
Forschungszentrum J\"{u}lich and JARA, 52425 J\"{u}lich, Germany}
\author{Maximilian Merte}
\affiliation{Peter Grunberg Institut and Institute for Advanced Simulation, 
Forschungszentrum J\"{u}lich and JARA, 52425 J\"{u}lich, Germany}
\affiliation{Institute of Physics, Johannes Gutenberg-University Mainz, 55128 Mainz, Germany}
\affiliation{Department of Physics, RWTH Aachen University, 52056 Aachen, Germany}
\author{Frank Freimuth}
\affiliation{Peter Grunberg Institut and Institute for Advanced Simulation, 
Forschungszentrum J\"{u}lich and JARA, 52425 J\"{u}lich, Germany}
\author{Stefan Bl\"{u}gel}
\affiliation{Peter Grunberg Institut and Institute for Advanced Simulation, 
Forschungszentrum J\"{u}lich and JARA, 52425 J\"{u}lich, Germany}
\author{Yuriy Mokrousov}
\affiliation{Peter Grunberg Institut and Institute for Advanced Simulation, 
Forschungszentrum J\"{u}lich and JARA, 52425 J\"{u}lich, Germany}
\affiliation{Institute of Physics, Johannes Gutenberg-University Mainz, 55128 Mainz, Germany}




\date{\today}                                         

\begin{abstract}
Motivated by the recently observed topological Hall effect in ultra-thin films of SrRuO$_3$ (SRO) grown on SrTiO$_3$ (STO) [001] substrate, we investigate the magnetic ground state and anomalous Hall response of the SRO ultra-thin films by virtue of spin density functional theory (DFT). Our findings reveal that in the monolayer limit of an SRO film, a large energy splitting of Ru-$t_{2g}$ states stabilizes an anti-ferromagnetic (AFM) insulating magnetic ground state. For the AFM ground state our Berry curvature calculations predict a large anomalous Hall response upon doping. From the systematic symmetry analysis we uncover that the large anomalous Hall effect arises due to a combination of broken time-reversal and crystal symmetries caused by the arrangement of non-magnetic atoms (Sr and O) in the SRO monolayer. We identify the emergent Hall effect as a clear manifestation of the so-called crystal Hall effect in terminology of \v{S}mejkal {\it et al.} arXiv:1901.00445 (2019),
and demonstrate that it persists at finite frequencies which is the manifestation of the crystal magneto-optical effect. Moreover, we find a colossal dependence of the AHE on the degree of crystal symmetry breaking also in ferromagnetic  SRO films, which all together points to an alternative explanation of the emergence of the topological Hall effect observed in this type of systems.
\end{abstract}

\maketitle {}
\section{Introduction} 

Owing to the remarkable thermal
properties of SrRuO$_3$ (SRO)\cite{thermal},
it's thin films and heterostructures are intensively investigated as a possible route to realize  oxide-based electronic devices \cite{spaldin1,ghosez}. Historically, ferromagnetic SRO is also one of the cornerstone materials in the field of the anomalous Hall (AHE) effect \cite{X1} and it plays an important role in modern spintronics~\cite{spintr1, spintr2}. 
Recently, the emergence of interface-stabilized skyrmions 
 was reported in SRO/SrIrO$_3$ heterostructures \cite{japan} via the measurements of the topological Hall effect in this system. By investigating the thickness dependence of SRO/SrIrO$_3$ bilayer, it was suggested that the skyrmion phase in this bilayer is driven by 
strong spin-orbit coupling of SrIrO$_3$,  which in combination with octahedral distortion leads to a sizeable Dzyaloshinskii-Moriya interaction at the interface thus leading to the formation of chiral structures. This finding triggered an immense activity aimed at the observation of skyrmions in SRO grown on SrTiO$_3$ (STO) $-$ a well known system which does not possess large intrinsic spin-orbit coupling.
Very recently, an evidence of the skyrmion phase deduced from the topological Hall effect measurements was reported in a thin film of SRO grown on STO \cite{china, korea}, however, the interpretation of these findings was questioned in several works\cite{silvia,X2}.
The existing controversy motivates a careful microscopic analysis of the AHE in SRO thin films from accurate first principles theory in order to gain a so far missing unambiguous understanding of the AHE by relating it to the structural properties of this exciting system. 


The anomalous Hall effect plays an important role in condensed matter physics and material science research owing in part to its intriguing quantum mechanical, relativistic and topological nature \cite{hall,nagaosa,von,sinova}. For ferromagnets, the Hall resistivity is expressed as $\rho_{xy}=R_{0}H+R_{s}M$, where $R_{0}$ and $R_{s}$ are ordinary and extraordinary Hall coefficients, and $H$, $M$ are the magnetic field and magnetization of the sample, respectively. In the latter expression the first term signifies the ordinary Hall contribution, while the second term is the spontaneous magnetization contribution, which later came to be known as the anomalous Hall effect. The strong deviations from the linear behavior of the AHE with the magnetic field, postulated above, are often interpreted nowadays as the fingerprints of formation of complex magnetic textures and serve as the markers of the formation of skyrmion order in materials which exhibit them\cite{top1,china, korea}.

Ferromagnets naturally lend themselves as the materials where the AHE is manifest owing to the fact that the presence of spin-orbit coupling (SOC) in combination with ferromagnetic magnetization results in broken time-reversal and spatial symmetries\cite{Mac} which are consistent with the AHE formation. 
On the other hand, antiferromagnetic (AFM) materials are attracting increasing attention owing to their prospects in the realm of AFM spintronics\cite{balt, jung,yura}.
While historically the subject of the AHE in non-collinear AFMs is a blossoming field intensively researched also nowadays\cite{kontani1,tanaka2,chen, kubler, naka,ajoy, zhang, sur,yura2}, the AHE has been assumed irrelevant in collinear AFMs, where the breaking of symmetry due to non-collinear magnetic order does not occur. Recently, the matter of AHE in collinear compensated AFMs has been pushed forward by \v{S}mejkal and co-workers\cite{libor}, who realized that the AHE in such materials can arise as a result of the 
structural symmetry breaking rather than the magnetic order itself. The finding of the so-called crystal Hall effect raises the question of the role that the AFM phases of structurally-complex materials play in the formation of measured anomalous Hall signal.


In this work we investigate the electronic and Hall transport properties of SRO thin films grown on STO $-$ a system which is reported to be an insulator in the ultrathin film limit \cite{toyo, chang, xia,prl}. 
By performing first principles density functional theory (DFT) calculations we find the ground state of SRO mono-layer film grown on STO [001] to be a compensated antiferromagnet 
with magnetic moment lies in the plane of the film in agreement with the experimental observation \cite{xia}.
Our Berry curvature calculations predict a large AHE in the compensated collinear AFM phase of SRO thin film when the Fermi energy lies outside of the electronic gap. While we find that the magnetic structure of thin film SRO alone generates no Hall response, from symmetry analysis we clearly identify that it is the lowering of structural symmetry associated with the octahedral distortion of the lattice of oxygen (O) and strontium (Sr) atoms that in combination with time-reversal symmetry breaking gives rise to the AHE in this system. The observed AHE thus falls under the category of the so-called crystal Hall effect, recently uncovered theoretically by \v{S}mejkal and co-authors~\cite{libor}, and we also show that it persists at finite frequencies resulting in the crystal magneto-optical effects. We further show that the octahedral distortion brings colossal modifications to the AHE of the ferromagnetic (FM) films as well.  We thereby suggest that the {\it crystal} part of the AHE in AFM and FM phases of SRO thin films can be an important ingredient for understanding the physics of topological Hall effect arising in this fascinating system.

\section{Computational Details}

DFT calculations were carried out with two different approaches: the full potential 
linearized augmented plane wave (FLAPW) method as implemented in the J\"{u}lich DFT code FLEUR, \cite{fleur}, and the plane-wave projected augmented wave (PAW) method as implemented in Vienna ab initio Simulation Package (VASP)\cite{vasp,paw}.
The structural optimization of the bulk as well as thin film structures was carried out using the VASP code maintaining the symmetry of the crystal. 
The positions of the atoms were relaxed towards equilibrium until the Hellman-Feynman forces became less than 0.001\,eV/\AA. 
The Monkhorst-Pack\cite{monk} $k$-point mesh of 8$\times$8$\times$6  and 10$\times$10$\times$10  was used for  structural optimization of bulk SRO and STO structures, respectively. This choice of the $k$-mesh and a plane-wave cutoff of 500\,eV were found to provide a good convergence of the total energy.

The Monkhorst-Pack $k$-point mesh of 8$\times$8$\times$2 was used for the structural optimization of SrO-terminated SRO thin films grown on STO. To realize the anti-ferromagnetic spin-configuration as well as to take into account the octahedral distortion of RuO$_6$ (both tilting and rotation) in-plane dimension of the simulation cell was fixed to $\sqrt{2}$ of the theoretically optimized lattice parameters of cubic STO [cf. Fig. 1(a)]. For the structural optimization of the thin film structure in the plane wave basis, we included 20\,\AA\, of vacuum to minimize the interaction between periodically repeated images along the $z$-axis. Then we carried out the structural optimization of thin film structure by relaxing the internal positions allowing for tilting and rotation of RuO$_6$ octahedra and keeping the in-plane lattice parameters fixed at $\sqrt{2}$ a$_{STO}$.

Using relaxed atomic positions of SRO monolayer, total energy calculations of different structures, the electronic structure calculations including the effect of spin-orbit coupling (SOC) and the AHE  calculations were carried out with the film version of the FLEUR code.\cite{fleur} For self-consistent calculations with the LAPW basis set a plane-wave cutoff of $k_{max}= 4.2$\,a.u.$^{-1}$ and the total of 576 $k$-points in the two-dimensional Brillouin zone were found to be sufficient for the convergence of the total energy. The plane wave cutoff for the potential (g$_{max}$) and exchange-correlation potential (g$_{max,xc}$) were set to 15.6 and 12.0\,a.u.$^{-1}$, respectively. 
The muffin-tin radii for Sr, Ru, O were set to 2.80\,\AA, 2.32\,\AA, and 1.31\,\AA, respectively. For calculations of the magnetic anisotropy energy the effect of SOC was included self-consistently using 48$\times$48 $k$-points in the two-dimensional Brillouin zone.

\begin{figure}[t!]
\begin{center}
\rotatebox{0}{\includegraphics[width=0.45\textwidth]{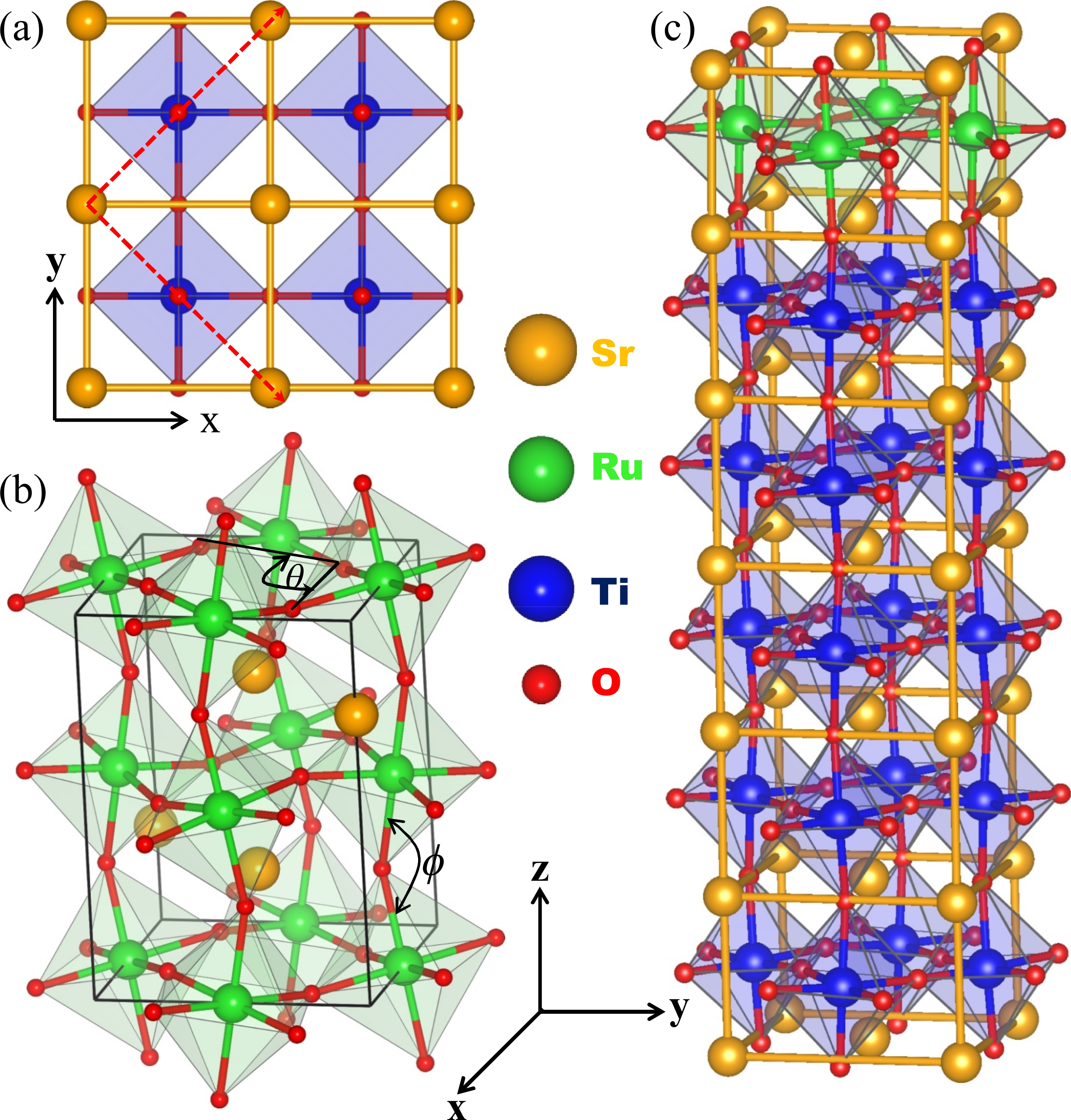}}
\end{center}
\caption{(a) Cubic crystal structure of bulk STO. Red arrows indicate the $\sqrt{2} \times \sqrt{2}$ supercell adopted to take into account the tilting and rotation of oxygen octahedra as well as the in-plane antiferromagnetic order of Ru moments. (b) Orthorhombic crystal structure of bulk SRO. Tilting angle, ($180^{0}-\phi$)/2, and rotation angle, ($90^{0}-\theta$)/2, of oxygen octahdera are  marked. (c) Thin-film structure of a monolayer of SRO grown on STO.}
\label{figure:1}
\end{figure}

We used the Perdew-Burke-Ernzerhof (PBE) \cite{pbe} exchange-correlation functional within the generalized gradient approximation (GGA). The electron-electron correlation effects beyond GGA at the magnetic Ru ions were taken into account by referring to the GGA+$U$ method\cite{lda+u}, where two key parameters $-$ the onsite Coulomb interaction strength $U$, and the  intra-atomic exchange interaction strength $J$ $-$ were computed using the constrained random phase approximation (cRPA) method \cite{crp1, crp2} as implemented in the SPEX code \cite{spex} using an $8\times 8\times1$ $k$-point grid,  resulting the values of $U=2.52$\,eV and $J=0.44$\,eV. 

\section{Structural properties}
Bulk SRO is found to be stabilized in an orthorhombic crystal structure below 850\,K with GdFeO$_3$-type distortion~\cite{koster} characterized by the tilting of the RuO$_6$ octahedra in alternate directions away from the $z$-axis and the rotation of the octahedra, as shown in Fig. 1(b). On the other hand, at room temperature, STO is known to exhibit perfect cubic structure without octahedral tilting and rotation \cite{sto}. We first optimize the lattice parameters of the bulk SRO and STO, keeping the symmetry of the structure fixed. We present in Table I the optimized lattice parameters for both structures, finding them to be in good agreement with previous studies $\cite{maha, scirep,zay}$. We find the distortion of RuO$_6$ octahedra which manifests in unequal bond lengths and deviations of O-Ru-O bond angles away from  90$^{\circ}$. We find optimized tilting angle, (180-$\phi$)/2, of 10.56$^\circ$ (corresponding to a Ru-O-Ru angle of 159$^\circ$), and rotation angle, (90-$\theta$)/2, of 7.56$^{\circ}$.

\begin{table}[t!]
\centering
\caption{\label{table3} Optimized GGA lattice parameters for bulk SrTiO$_3$ and SrRuO$_3$. For comparison, experimental values are also shown (marked as ``Exp."). }
\begin{tabular}{c|@{\hspace{0.2cm}}c@{\hspace{0.5cm}}c@{\hspace{0.8cm}}c@{\hspace{0.8cm}}c@{\hspace{0.2 cm}}}
\hline\hline
system & Type& a[ \AA ] & b[ \AA ] & c[ \AA ] \\  \hline
cubic SrTiO$_3$ & Exp.\cite{sto}& 3.91&3.91&3.91 \\
&GGA & 3.957 & 3.957 & 3.957  \\  
\hline
orthorombic SrRuO$_3$&Exp.\cite{sro}& 5.567 & 5.530 & 7.845  \\
&GGA & 5.628 & 5.616 & 7.957  \\         
\hline\hline
\end{tabular}
\end{table}

 
Due to the lattice mismatch between orthorhombic SRO and cubic STO, a thin film of SRO grown on STO substrate corresponds to 0.47$\%$ of compressive strain.
In the optimized structure of SRO thin film, we observe a marked change in the RuO$_6$ octahedral distortion in terms of  bond lengths and Ru-O-Ru bond angles,  making RuO$_6$ octahedra much more distorted as compared to those of bulk RuO$_6$ octahedra. 
The rotation angle of RuO$_6$ octahedra is found to be increased by 5.84$^{\circ}$ as compared to that of the bulk structure. These changes 
in the octahedral distortion 
have a crucial impact on the energetic position of Ru-t$_{2g}$ states, thus directly influencing the electronic structure, as discussed later.

\section{Electronic structure of SRO films}
First, with the GGA+$U$ method, we compare the total energies of different magnetic structures $-$ non-magnetic, ferromagnetic and antiferromagnetic $-$ in order to find out the magnetic ground state of SRO films. Considering the calculated value of $U=2.52$\,eV and $J=0.44$\,eV at the Ru site, the AFM  state is found to be more stable as compared to the ferromagnetic one, by 66\,meV per Ru atom.  In Fig. 2(a) we show the orbitally-resolved density of states (DOS) as calculated with GGA+$U$, assuming the AFM spin structure. The DOS clearly shows the insulating nature of the ground state  in agreement with the previously reported DFT+$U$ and DMFT+$U$ results~\cite{maha, scirep,dmft}, as well as with recent experimental data consistent with the emergence of an insulating state with no net moment \cite{chang,xia,prl}.

The GGA+$U$ spin moment at the Ru ion is found to be 1.26\,$\mu_B$, which is consistent with the low spin state of Ru$^{4+}$ ion ($d^{4}$:t$^{3}_{2g \uparrow}$,t$^{1}_{2g \downarrow}$). 
The $d$ states of Ru are exchange and crystal-field split. As evident from Fig. 2, Ru-$d_{xy}$ state becomes completely occupied in both spin-channels while $d_{xz,yz}$ states are filled in the majority spin channel and the minority spin channel remains empty. This gives rise to an insulating state with a gap in both spin channels. From the calculated GGA+$U$ DOS  at $T = 0$\,K the value of the gap is found to be $~$1.1\,eV, which is in a good agreement to the DMFT result \cite{dmft}.

\begin{figure}[t!]
\begin{center}
\rotatebox{0}{\includegraphics[width=0.48\textwidth]{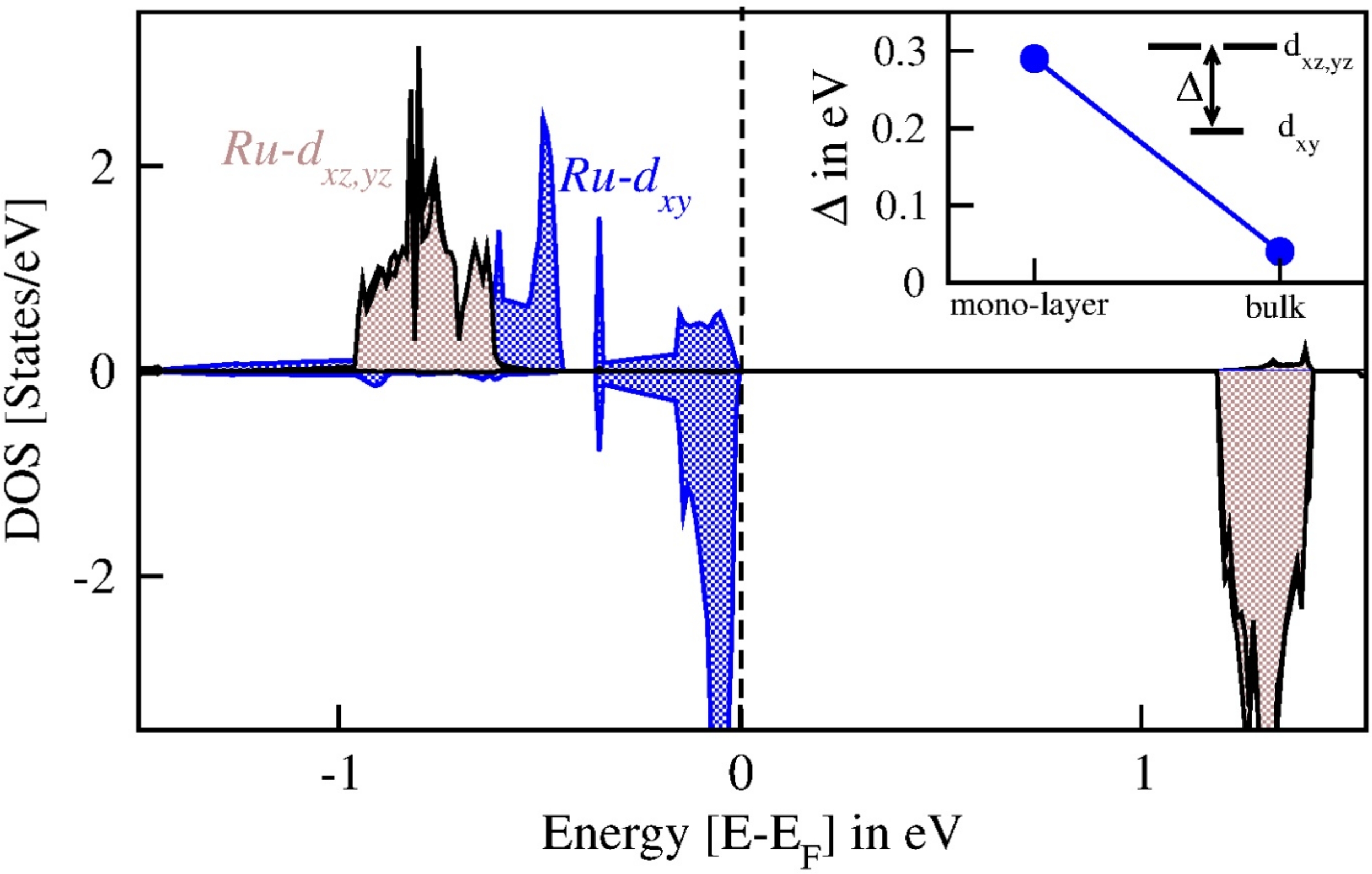}}
\end{center}
\caption{Spin-polarized density of states as computed in GGA+$U$, projected onto the octahedral crystal-field split Ru-t$_{2g}$ states.  The
distortion of RuO$_6$ octahedra gives rise to a large splitting $\Delta= \epsilon_{xy}-\epsilon_{xz,yz}$ between the  $d_{xy}$ and $d_{xz,yz}$ states. Inset shows the variation of $\Delta$ in bulk orthorhombic SRO and in the monolayer of SRO grown on STO.}
\label{figure:2}
\end{figure}

To gain the microscopic insight into the formation of the AFM insulating state of the SRO monolayer, we compute the energy-level diagram of Ru-$d$ states, employing the technique of maximally localized Wannier functions (MLWFs)\cite{wan1, wan2, v3citation,freimuth}, considering only the 
Ru-$d$ Hamiltonian constructed out of non-spin-polarized GGA calculations. With the information of the energy level position of Ru-$d$ states, obtained from the real-space representation of the Hamiltonian in the MLWFs basis, we calculate the energy level difference $\Delta$=$\epsilon_{yz,xz}$ - $\epsilon_{xy}$ between Ru-$d_{yz,xz}$ and $d_{xy}$ states. The calculated energy level difference $\Delta$ for the SRO monolayer and for bulk is shown in the inset of  Fig. 2. A larger $\Delta$ of 0.3\,eV is observed in the monolayer limit due to the octahedral distortion of RuO$_6$ in terms of unequal bond lengths. The large energy separation between $\epsilon_{xz,yz}$ and $\epsilon_{xy}$ roughly corresponds to the half-filled ($d_{xz,yz}$ states) two-band case within the ionic picture of Ru$^{4+}$, as observed e.g. in Ca$_2$RuO$_4$ \cite{lie}. The two-band half-filled case naturally stabilizes the AFM spin ordering by Ru-O-Ru super-exchange interaction\cite{khom}.


\section{Magnetocrystalline anisotropy energy}
Next, we investigate the magnetic anisotropy energy (MAE) of SRO mono-layer. We obtain the MAE from the difference in total energy of the AFM state with spins aligned with the crystal axes $y$ and $z$, with the total energy of the system with staggered magnetization along the $x$-axis, showing the results in Table II. 
We find $x$-axis to be the easy axis of the system. Despite the fact that the two in-plane lattice constants of the SRO thin film are equal, a small energy difference between the $x$ and $y$ directions of the staggered magnetization of about 0.05 meV/Ru is observed, which can be traced back to the orthorhombic distortion of RuO$_6$ cages. Notably, the value of MAE of about 7.31\,meV/Ru is quite large. 


\begin{table}[t!]
\centering
\caption{\label{table2} Magnetocrystalline anisotropy energy in meV/Ru and the Ru-site projected spin moment (M$_s$) and orbital moment M$_{L}$ for monolayer  SRO grown on STO. The data are presented considering different crystallographic directions of the staggered magnetization.}
\begin{tabular}{c|@{\hspace{0.1cm}}c@{\hspace{0.4cm}}c@{\hspace{0.4cm}}c@{\hspace{0.1cm}}}
\hline\hline
 
&[100]&[010]&[001] \\ \hline
MAE(meV/Ru)&0.000&0.055&7.308 \\
M$_{s}$ ($\mu_{B}$)&1.251&1.251&1.254 \\
M$_{L}$ ($\mu_{B}$)&0.141&0.139&0.001 \\  
\hline\hline
\end{tabular}
\end{table}

In Table II, we also show the calculated local Ru spin and orbital magnetic moments for different directions of the staggered magnetization. For all three directions the orbital moments of Ru$^{2+}$($d^{4}$) are of the same sign as the spin moments, which is expected due to  more than half-filled Ru-t$_{2g}$ sub-shell. Interestingly, the value of orbital moment is found to be two orders of magnitude smaller for the staggered magnetization along $z$  as compared to two other crystallographic directions. Referring to the Bruno's interpretation of MAE~\cite{Bruno}, such a suppression of the orbital moment for the out-of-plane magnetization energetically promotes the in-plane direction of the magnetization (in agreement with experimental observation \cite{xia}), given a small variation in spin moment. 


\begin{figure*}[t!]
\begin{center}
\rotatebox{0}{\includegraphics[width=0.98\textwidth]{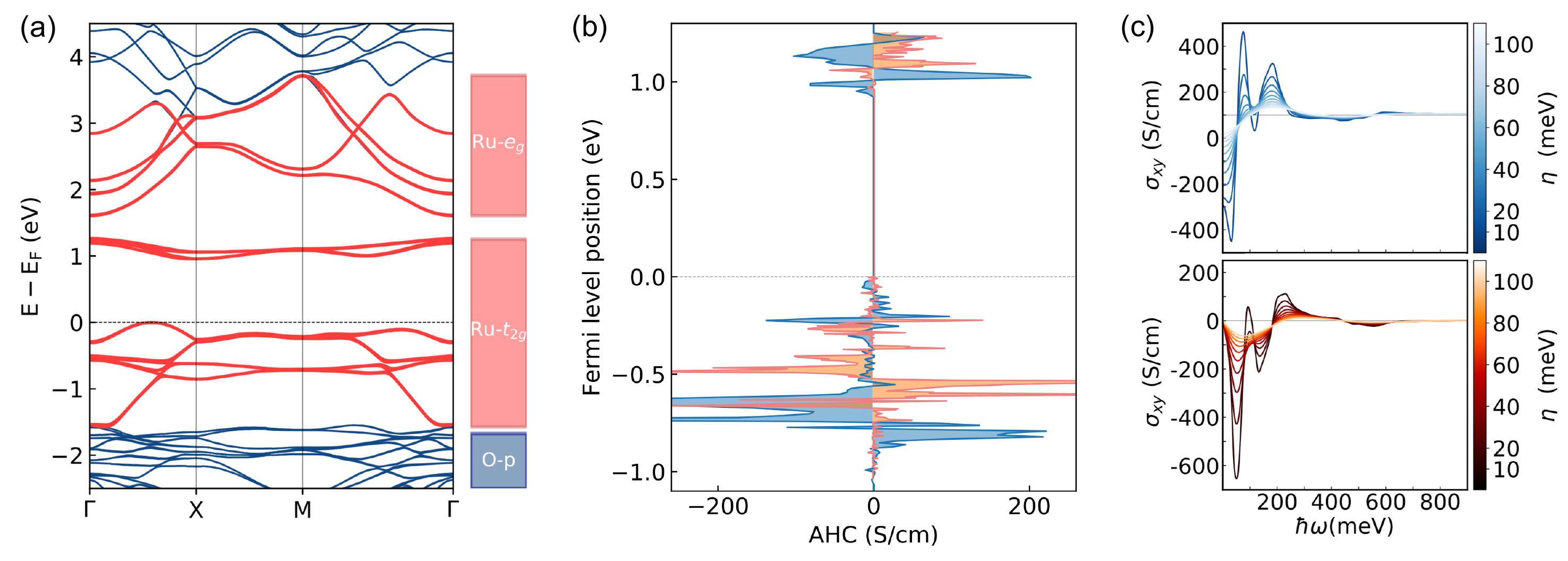}}
\end{center}
\caption{ (a) Band structure of SRO monolayer grown on STO in the AFM  state with staggered magnetization along the $x$-axis, including the effect of SOC. Blue lines: GGA+U+SOC(100) first principles electronic bands. Red lines: Wannier-interpolated band structure. 
The dominant orbital character of the states is shown on the side.
(b) Corresponding computed AHC as a function of Fermi level position. The orange line reflects the AHC considering the rotation of RuO$_6$ octahedra only. (c) Real (blue) and imaginary (red) part of  the magneto-optical conductivity as a function of photon energy. The Fermi level was set to $E_F=-0.6$ eV, smearing values of 10 to 100 meV were selected and are indicated by the color coding.     
}
\label{figure:4}
\end{figure*}

\section{Anomalous Hall and magneto-optical conductivity}
Having understood the origin of the magnetic ground state and the magneto-crystalline anisotropy energy, next we proceed to investigate the anomalous Hall effect of the SRO monolayer, motivated by the recently observed anomalies in the behavior of the AHE as a function of applied magnetic field in this system\cite{china, korea}. 
Here, we assess the intrinsic Berry curvature contribution to the AHE employing the  Wannier interpolation technique\cite{wang}. To compute the Berry curvature, we first construct a tight-binding MLWFs Hamiltonian 
projected from the GGA+$U$+SOC(100) Bloch wavefunctions. Atomic-orbital-like MLWFs of Ru-t$_{2g}$ and e$_{g}$ states are considered to construct the minimal tight-binding Hamiltonian, which reproduces the spectrum of the system in a limited energy window around the Fermi energy. From this Hamiltonian the Berry curvature is calculated according to 
\begin{eqnarray}
\begin{aligned}
\Omega_{n}(\mathbf{k}) = -\hslash^{2} \sum_{n \neq m} \frac{\operatorname{2 Im} \langle u_{n\mathbf{k}}| \hat{ v}_{x}|u_{m\mathbf{k}}\rangle \langle u_{n\mathbf{k}}|\hat{v}_{y}|u_{m\mathbf{k}}\rangle}{(\epsilon_{n\mathbf{k}}-\epsilon_{m\mathbf{k}})^{2}},
\end{aligned}
\end{eqnarray}
 where $\Omega_{n}(\mathbf{k})$ is the Berry curvature of band $n$, $\hslash\hat{v}_{i} ={\partial \hat{H}(\mathbf{k})}/{\partial k_{i}} $ is the $i$'th velocity operator, $u_{n\mathbf{k}}$ and $\epsilon_{n\mathbf{k}}$ are the eigenstates and eigenvalues of the Hamiltonian $\hat{H}(\mathbf{k})$, respectively.

In Fig. 3(a) we show the comparison of the {\it ab initio} GGA+$U$+SOC(100) band structure of SRO monolayer (blue lines) with that obtained by diagonalization of Ru-$d$ projected Wannier Hamiltonian (red lines), finding an excellent agreement in the region of $\pm2$\,eV with respect to the middle of the gap. 
From this Hamiltonian we calculate the Berry curvature on a $50\times 50$ $k$-mesh employing an adaptive $5\times 5$ refinement scheme\cite{yao} at points where the value of the Berry curvature exceeds 50\,a.u. These numerical parameters provide well-converged values of the  anomalous Hall conductivity (AHC) determined as 
\begin{eqnarray}
\begin{aligned}
\sigma_{x y}=& -\hbar e^{2} \int_{BZ} \frac{d^{3} k}{(2 \pi)^{3}}  \Omega(\mathbf{k}),
\end{aligned}
\end{eqnarray}
where $\Omega(\mathbf{k})$ is the sum (for each k) of Berry curvatures over the occupied bands.
For some values of the Fermi energy we checked that our AHC values are stable with respect to the choice of the MLWFs reproducing the band structure in the whole energy window of occupied states.
Our calculations of the AHC in the AFM monolayer SRO are shown in Fig. 3(b) as a function of the Fermi energy. Notably, we find that while the gapped system is a topologically-trivial insulator (since the quantized value of the AHC is zero in the gap), away from the gap  a large contribution to the AHC $-$ comparable to that observed in such elemental ferromagnets as hcp Co~\cite{hcp-co} $-$ emerges. This seems counter-intuitive given the compensated AFM nature of this two-dimensional material exhibiting an in-plane direction of staggered magnetization.


To understand the microscopic origin of the observed large Hall response, we consider the SRO monolayer without taking into account the octahedral rotation and titling. While in this case our GGA+$U$+SOC calculations also predict the AFM insulating ground state with an easy axis along the $x$ direction, the computed Hall response is found to be zero within the accuracy of AHC calculations. This makes it clear that it is the structural lowering of symmetry due to octahedral tilting and rotation, which is responsible for the AHE in the studied material. 
To separately sort out the impact of the octahedral tilting and rotation on the AHC, we start with an undistorted  SRO monolayer with zero AHC and introduce an octahedral rotation of 7.56$^{\circ}$. The corresponding calculated AHC, shown in Fig. 3(b) with a orange line, differs significantly from that  computed including rotation and tilting at the same time, which means that both channels for symmetry lowering are equally important in giving rise to the large AHE in monolayer SRO.


The role of octahedral distortion for the AHE is ultimately reflected in the symmetry breaking that it causes. We thus analyze the symmetries of the two structures to gain a better insight into the microscopic origin of the Hall response. In Fig. 4 we present side to side the top view of the monolayer without (a) and with (b) tilting and rotation of the octahedra.  Clearly, the atomic structure in Fig. 4(a) has four-fold rotational symmetry (C$_{4}$) as well as four reflection symmetries with respect to two mirror plane accommodating $x$- and $y$-axes (m$_{x}$ and m$_{y}$), and two mirror planes accommodating the diagonals (m$_{d}$ and m$_{d'}$). For this structure, applying the time-reversal ($\tau$) symmetry operation followed by a translation by half a lattice constant (t$_{1/2}$) one recovers the same magnetic structure. This ultimately results in zero net AHC in the latter case.

In contrast, the octahedral tilting and rotation present in the structure of Fig. 4(b) caused by the asymmetric position of the nonmagnetic atoms oxygen (O) and strontium (Sr), breaks all reflection symmetries as well as the C$_{4}$ symmetry. Correspondingly, applying the t$_{1/2}\tau$ operation one arrives at a crystal with opposite structural chirality. As realized by \v{S}mejkal and co-workers, this serves as the ultimate reason for the emergence of the AHE in case of a structure from Fig. 4(b). Given that the job of symmetry breaking necessary for the AHE is done here by the cage of non-magnetic atoms, the AHE in this context has been coined as the crystal Hall effect~\cite{libor}. Our calculations thus mark the emergence of the crystal Hall effect for the AFM state of the thin films of SRO.


The conclusions of the symmetry analysis that we performed should hold not only for the anomalous Hall effect, but generally for magneto-optical effects, as the two classes of phenomena have the same symmetry properties. To demonstrate this, we consider the magneto-optical (MO) conductivity, which is the extension of the d.c.~AHC that we discussed previously, to the case of an electric field of finite frequency $\omega$.   
The MO conductivity was calculated using the Kubo expression \cite{yao}
\begin{equation}
\begin{aligned}
\sigma_{x y}(\omega)=& \hbar e^{2} \int \frac{d^{3} k}{(2 \pi)^{3}} \sum_{n \neq m}\left(f_{n \mathbf{k}}-f_{m \mathbf{k}}\right) \\
& \times \frac{\operatorname{Im}\left[\left\langle u_{n \mathbf{k}}\left|\hat{v}_{x}\right| u_{m \mathbf{k}}\right\rangle\left\langle u_{m \mathbf{k}}\left|\hat{v}_{y}\right| u_{n \mathbf{k}}\right\rangle\right]}{\left(\epsilon_{n \mathbf{k}}-\epsilon_{m \mathbf{k}}\right)^{2}-(\hbar \omega+i \eta)^{2}},
\end{aligned}
\end{equation}
where   
$f_{n \mathbf{k}}$ is the Fermi-Dirac distribution function, $\hbar\omega$ is the photon energy and $\eta$ is a smearing parameter. 
In Fig. 3(c) we show the real and imaginary parts of the MO conductivity for the AFM state of SRO monolayer, considering octahedral rotation and titling, as a function of photon energy, while positioning the Fermi energy in the middle of Ru-t$_{2g}$ states at $-0.6$\,eV. Calculations with smearing parameters ranging from 10 to 100\,meV  were performed. The smearing of 
10\,meV corresponds to the ``clean" case of a perfect crystal and corresponsingly the zero-frequency limit of the real part of the MO conductivity coincides nicely with the d.c. value of the AHC at that energy.
While increasing the value of the smearing parameter $-$ which qualitatively corresponds to the inverse of the quasi-particle lifetime of the electronic states determined by the degree of disorder in the film $-$ eventually suppresses the overall magnitude of the MO conductivity and features in its $\omega$-dependence, it exhibits a complex structure and a very sizeable magnitude in its real and complex parts over a large range of $\eta$. This marks the emergence of {\it crystal} magneto-optical effects in our system, which can be also probed with magneto-optical experimental techniques via the measurements of Kerr and Faraday effects.  


\begin{figure}[t!]
\begin{center}
\rotatebox{0}{\includegraphics [width=0.45\textwidth]{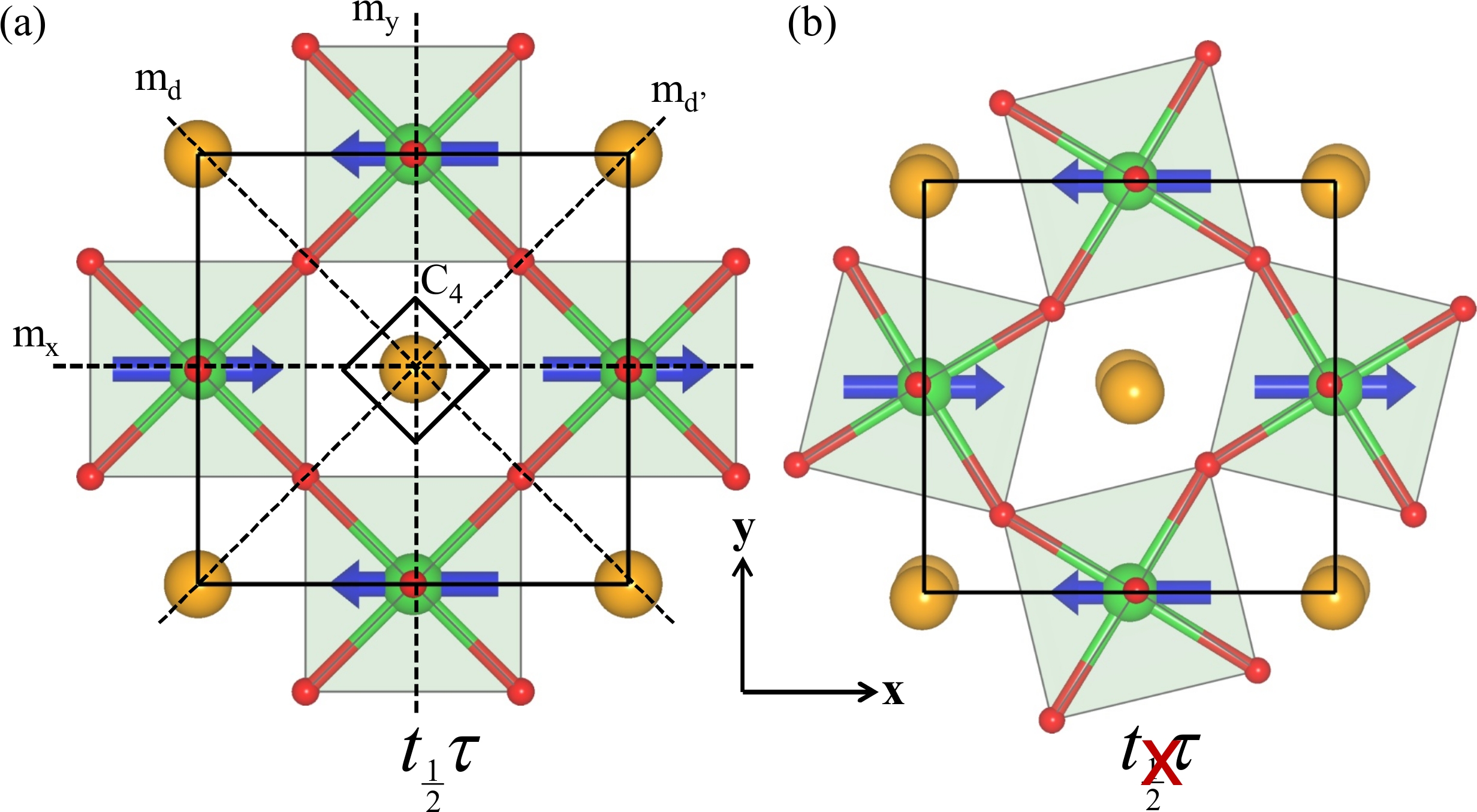}}
\end{center}
\caption{(a) Top view of the monolayer SRO in absence of octahedral tilting and rotation.
 4-fold rotation (C$_4$) axis, two mirror planes (m$_{x}$, m$_{y}$) along $x$ and $y$  axes, and two diagonal mirror planes m$_{d}$, m$_{d'}$ are marked. A combination of time-reversal symmetry  with a translation by half a lattice constant, t$_{1/2}\tau$, is preserved in this case. (b) Top view of the monolayer SRO taking into account the octahedral tilting and rotation. Breaking of the t$_{1/2}\tau$ symmetry by the nonmagnetic atoms (oxygen and strontium) results in the emergence of the AHE in this case.}
\label{figure:4}
\end{figure}

\section{Discussion}

In this work, by performing first principles calculations we predict that doped SRO monolayer grown on STO will exhibit a strong AHE and magneto-optical effects in its AFM ground state. This finding has consequences which are two-fold. Firstly, we can translate our calculations into a prediction that  in ultra-thin metallic films of SRO, which were reported to be AFM in a range of thickness from mono- to bi-layers~\cite{prl}, the AHE and MO effects can be observed experimentally  despite the AFM ground state. This roots in the understanding that the physics of the AHE in the case of larger SRO thickness will be  governed by the same microscopic mechanism of octahedral distortion.
The absence of the experimental signatures of the AHE in the latter case can hint at presence of structural domains where the sense of the octahedral distrotion $-$ i.e. structural chirality $-$ is opposite, since the crystal AHE exhibited by SRO thin films switches sign upon switching the structural chirality.  
\begin{figure}[t!]
\begin{center}
\rotatebox{0}{\includegraphics [width=0.45\textwidth]{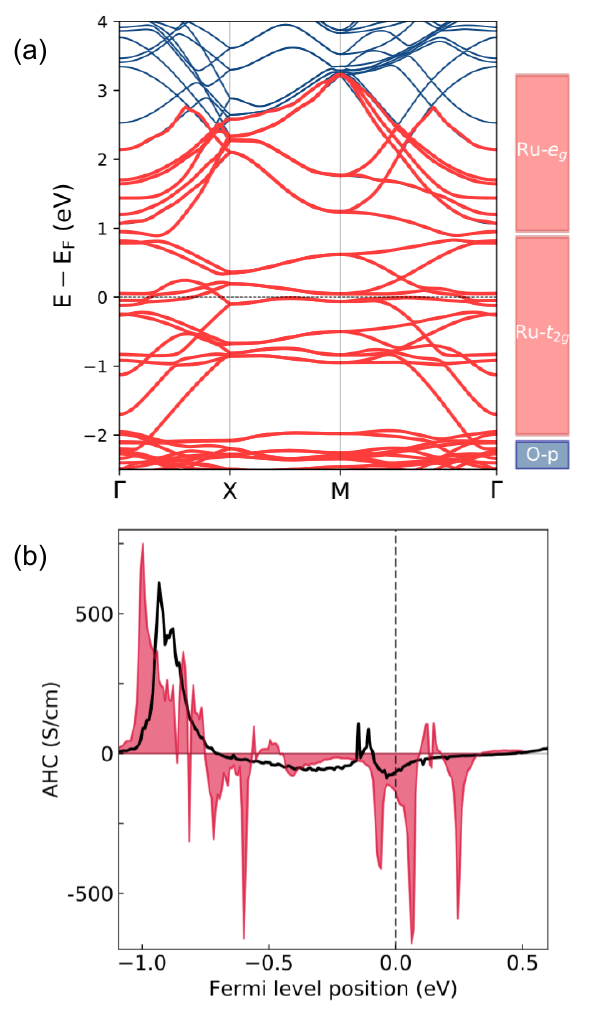}}
\end{center}
\caption{(a) Band structure of SRO monolayer grown on STO in the FM state with magnetization along the $z$-axis, including the effect of SOC. Blue lines: GGA+SOC(001) first principles electronic bands. Red lines: Wannier-interpolated Ru-d+O-p band structure. 
The dominant orbital character of the states is shown on the side.
(b) Corresponding computed AHC as a function of Fermi level position. The shaded area and black lines reflect the AHC with and without rotation and tilting of RuO$_6$ octahedra, respectively.}
\label{figure:5}
\end{figure}
This brings us to the second point: the  interplay of structural properties with the AHE and MO effects in thin {\it ferromagnetic} (FM) films of SRO. 
Namely, our findings of the large crystal AHE in the AFM phase point intuitively at the strong dependence of the AHE in the FM system on the sense of structural chirality and degree of octahedral distortion. To demonstrate this point, we perform calculations of the AHE for the FM state of the SRO monolayer. We relax the structure and from total energy calculations find that the easy axis of the system is the $z$-axis. Analogously to the case above, we used Wannier interpolation technique to interpolate the electronic structure (see Fig.\,5(a)) of the system, and compute the AHC. The calculations of the AHE, presented in Fig.\,5(b), reveal a remarkable influence of the modest octahedral distortion $-$ that we find to minimize to total energy of the system $-$ on the AHC: as compared to the case where the distortion is not taken into account, the magnitude and structure of the AHC as a function of the band filling are drastically modified in the vicinity of the Fermi energy when the distortion is taken into account for this metallic system. Since this effect is originated in the redistribution of the bands around the Fermi energy driven by symmetry lowering, we are confident that the ``crystal" sensitivity of the AHE is prominent in thicker FM films as well.
Given the intricate structural properties of deposited thin films of SRO ~\cite{china,korea,Zhang}, we expect a very complex distribution of tilting and rotation angles of RuO$_6$ octahedra as a function of distance from the surface or interface with another oxide. Based on our calculations, we dare suggest that the structural chirality thus presents an additional powerful variable in the complex physics of the AHE in SRO films and their interfaces, which so far has not been seriously considered as an active player in directly determining the transport properties of latter materials. We  speculate that the physics of structural reorientation and chiral phase formation via its influence on the AHE in FM and AFM thin SRO films can serve as a plausible explanation for  complex behavior of the AHE in this family as a function of the temperature, thickness and an external magnetic field, reported in several studies~\cite{china, korea,korea2}, and which is commonly interpreted as topological Hall effect due to formation of chiral spin structures. Pursuing this line of thought presents an exciting venue for future experimental and theoretical research.
\section{Acknowledgements}
We acknowledge extensive discussions with Libor \v{S}mejkal, Ionela Lindfors-Vrejoiu, Jan-Philipp Hanke, Jairo Sinova and Thomas Lorentz. We acknowledge funding from Deutsche For\-schungs\-gemeinschaft (DFG) through SPP 2137  ``Skyrmionics", the Collaborative Research Center SFB 1238, and project MO 1731/5-1. The work was funded also by the Deutsche Forschungsgemeinschaft (DFG, German Research Foundation) - TRR 173 - 268565370. Y.M. and S.B.\ acknowledge the DARPA TEE program through grant MIPR\# HR0011831554 from DOI.   Simulations were performed with computing resources granted by JARA-HPC  from RWTH Aachen University and Forschungszentrum J\"ulich under projects jiff38, jiff40 and jpgi11.


\begin{thebibliography}{100}
\bibitem{thermal}H. N. Lee, H. M. Christen, M. F. Chisholm, C. M. Rouleau, and D. H. Lowndes, Thermal stability of epitaxial SrRuO$_3$ films as a function of oxygen pressure, Appl. Phys. Lett. {\bf 84}, 4107 (2004).

\bibitem{ghosez} J. Junquera and P. Ghosez, Critical thickness for ferroelectricity in perovskite ultrathin films, Nature {\bf 422}, 506 (2003).
\bibitem{spaldin1}M.  Stengel  and  N.  A.  Spaldin, Origin of the dielectric dead layer in nanoscale capacitors, Nature  {\bf 443},  679 (2006). 


\bibitem{X1} Z. Fang, \textit{et al.} The Anomalous Hall Effect and Magnetic Monopoles in Momentum Space, Science {\bf 302}, 5642 (2003).
 
\bibitem{spintr1} I. Žutić, J. Fabian, and S. D. Sarma, Spintronics: Fundamentals and applications, Rev. Mod. Phys. {\bf 76}, 323 (2005).
\bibitem{spintr2} D. Awschalom and M. Flatte, Challenges for semiconductor spintronics, Nat. Phys. {\bf 3}, 153 (2007).



\bibitem{japan} J. Matsuno,  \textit{et al.} Interface driven topological Hall effect in SrRuO$_3$-SrIrO$_3$ bilayer, Sci. Adv. {\bf 2}, e1600304 (2016).



\bibitem{china} Y. Gu \textit{et al.} Interfacial oxygen-octahedral-tilting-driven electrically tunable topological Hall effect in ultrathin SrRuO$_3$ films, J. Phys. D: Appl. Phys. {\bf 52}, 404001 (2019).
\bibitem{korea}B. Sohn, \textit{et al.} Emergence of
robust 2D skyrmions in SrRuO$_3$ ultrathin film without the capping layer, arXiv 1810.01615 (2018).
\bibitem{silvia}D. J. Groenendijk, \textit{et al.} Berry phase engineering at oxide interfaces, arXiv:1810.05619 (2018).
\bibitem{X2}G. Malsch, \textit{at al.} Correlating the nanoscale structural, magnetic and magneto-transport properties in SrRuO$_3$-based perovskite oxide ultra-thin films, arXiv:1910.01474 (2019).


\bibitem{hall} E. H. Hall, On a new action of the magnet on electric currents, Am. J. Math. {\bf 2}, 287 (1879). 
\bibitem{nagaosa} N. Nagaosa, J. Sinova, S. Onoda, A. H. MacDonald, and N. P. Ong, Anomalous Hall effect, Rev. Mod. Phys. {\bf 82},1539-1592 (2010). 
\bibitem{von} K. von Klitzing, The quantized Hall effect, Rev. Mod. Phys. {\bf 58}, 519 (1986). 
\bibitem{sinova} J. Sinova, S. O. Valenzuela, J. Wunderlich, C. H. Back, and T. Jungwirth, Spin Hall effects, Rev. Mod. Phys. {\bf  87}, 1213 (2015).

\bibitem{top1} A. Neubauer, \textit{et al.} Topological Hall effect in the A phase of MnSi, Phys. Rev. Lett. {\bf 102}, 186602 (2009).



\bibitem{Mac} Y. MacHida, S. Nakatsuji, S. Onoda, T. Tayama, and T. Sakakibara, Time-reversal symmetry breaking and
spontaneous Hall effect without magnetic dipole order, Nature {\bf 463}, 210-213 (2010).
\bibitem{balt} V. Baltz, A. Manchon, M. Tsoi, T. Moriyama, T. Ono, and Y. Tserkovnyak, Antiferromagnetic spintronics, Rev. Mod. Phys. {\bf 90}, 015005 (2018).
\bibitem{jung} T. Jungwirth, X. Marti, P. Wadley, and J. Wunderlich, Antiferromagnetic spintronic, Nat. Nanotech.  {\bf 11}, 231 (2016).
\bibitem{yura} L. Šmejkal, Y. Mokrousov, B. Yan, and A. H. MacDonald, Topological antiferromagnetic spintronics, Nat. Phys.{ \bf 14}, 242 (2018).


\bibitem{kontani1}T. Tomizawa and H. Kontani, Anomalous Hall effect due to noncollinearity in pyrochlore compounds: Role of orbital Aharonov-Bohm effect, Phys. Rev. B {\bf 82}, 104412 (2010).

\bibitem{tanaka2}T. Tomizawa and H. Kontani, Anomalous Hall effect in the t$_{2g}$ orbital kagome lattice due to noncollinearity: Significance of the orbital Aharonov-Bohm effect, Phys. Rev. B {\bf 80}, 100401(R) (2009).

\bibitem{chen} H. Chen, Q. Niu, and A. H. MacDonald, Anomalous Hall effect arising from noncollinear antiferromagnetism, Phys. Rev. Lett. {\bf 112}, 017205 (2014).
\bibitem{kubler}J. K\"ubler and C. Felser, Non-collinear antiferromagnets and the anomalous Hall effect, EPL {\bf 108}, 67001 (2014).
\bibitem{naka} S. Nakatsuji, N. Kiyohara, and T. Higo, Large anomalous Hall
effect in a non-collinear antiferromagnet at room temperature, Nature {\bf 527}, 212 (2015).
\bibitem{ajoy} A. K. Nayak, J. E. Fischer, Y. Sun, B. Yan, J. Karel, A. C. Komarek, C. Shekhar, N. Kumar, W. Schnelle, J. Kübler, C.
Felser, and S. P. P. Parkin, Large anomalous Hall effect driven by a nonvanishing Berry curvature in the noncollinear antiferromagnet Mn$_3$Ge, Sci. Adv. { \bf 2}, e1501870 (2016).
\bibitem{zhang} Y. Zhang, Y. Sun, H. Yang, J. Železný, S. P. P. Parkin, C. Felser,
and B. Yan, Strong anisotropic anomalous Hall effect and spin Hall effect in the chiral antiferromagnetic compounds Mn$_3$X
(X=Ge, Sn, Ga, Ir, Rh, and Pt), Phys. Rev. B { \bf 95}, 075128 (2017).
\bibitem{sur} C. S\"urgers, G. Fischer, P. Winkel, and H. V.  L\"ohneysen, H. V.
Large topological Hall effect in the non-collinear phase of an antiferromagnet, Nat. Commun. {\bf 5}, 3400 (2014).
\bibitem{yura2} X. Zhou, J.-P. Hanke, W. Feng, F. Li, G.-Y. Guo, Y. Yao, S. Bl\"ugel, and Y. Mokrousov, Spin-order dependent anomalous
Hall effect and magneto-optical effect in the noncollinear antiferromagnets Mn$_3$XN with X = Ga, Zn, Ag, or Ni, Phys. Rev. B {\bf 99}, 104428 (2019).
\bibitem{libor} L. Šmejkal, R. González-Hernández, T. Jungwirth, and J. Sinova, Crystal Hall effect in collinear antiferromagnets,
arXiv:1901.00445 (2019).

\bibitem{prl}S. G. Jeong, \textit{et al.} Phase Instability amid Dimensional Crossover in Artificial Oxide Crystal, Phy. Rev. Lett. {\bf 124}, 026401 (2020).
\bibitem{toyo} D. Toyota, \textit{et al.} Thickness-dependent electronic structure of ultrathin SrRuO$_3$ films studied by in situ photoemission spectroscopy, Appl. Phys. Lett. { \bf 87}, 162508 (2005).
\bibitem{chang} Y. J. chang, \textit{et al.} Fundamental thickness limit of itinerant ferromagnetic SrRuO$_3$ thin films, Phys. Rev. Lett. {\bf 103}, 057201 (2009).
\bibitem{xia} J. Xia, W. Siemons, G. Koster, M. R. Beasley, A. Kapitulnik, Critical thickness for itinerant ferromagnetism in ultrathin films of SrRuO$_3$, Phys. Rev. B { \bf 79}, 140407(R) (2009).
\bibitem{fleur} www.flapw.de 
\bibitem{vasp} G. Kresse, and D. Joubert, From ultrasoft pseudopotentials to the projector augmented-wave method, Phys. Rev. B  { \bf 59}, 1758-1775 (1999).
\bibitem{paw} P. E Bl\"ochl, Projector augmented-wave method, Phys. Rev. B {\bf 50}, 17953-17979 (1994).
\bibitem{monk} H. J. Monkhorst and J. D. Pack, Special points for Brillouin-zone integrations, Phys. Rev. B {\bf 13}, 5188 (1976).
\bibitem{pbe} J. P. Perdew, K. Burke, M.  Ernzerhof, Generalized gradient approximation made simple, Phys. Rev. Lett. {\bf 77}, 3865-3868 (1996).
\bibitem{lda+u} V. I. Anisimovdag, F. Aryasetiawanddag and A. I. Lichtenstein, First-principles calculations of the electronic structure and spectra of strongly correlated systems: the LDA+ U method, J. Phys.: Condens. Matter {\bf 9} 767 (1997).
\bibitem{crp1} F. Aryasetiawan, M. Imada, A. Georges, G. Kotliar, S. Biermann, and A. I. Lichtenstein, Frequency-dependent local interactions and low-energy effective models from electronic structure calculations, Phys. Rev. B { \bf 70}, 195104 (2004).
\bibitem{crp2} E. Şaşıoğlu, C. Friedrich, and S. Bl\"ugel, Strength of the Effective Coulomb Interaction at Metal and Insulator Surfaces, Phys. Rev. Lett.  { \bf 109}, 146401 (2012).
\bibitem{spex} C. Friedrich, S. Blügel, and A. Schindlmayr, Efficient implementation of the GW approximation within the all-electron FLAPW method, Phys. Rev. B { \bf 81}, 125102 (2010).
\bibitem{koster} G. Koster, \textit{et al.} Structure, physical properties, and applications of SrRuO$_3$ thin films, Rev. Mod. Phys. { \bf 84}, 253-298 (2012).

\bibitem{sto} A. Leonarska, K. Szot, A. Ratuszna, Temperature evolution of the crystal structure in SrTiO$_3$ doped by W$^{6+}$, Ni$^{3+}$, Fe$^{3+}$ and La$^{3+}$, Phase Transitions {\bf 84}, 1015-1027 (2011).


\bibitem{maha} P. Mahadevan, F. Aryasetiawan, A. Janotti,  T. Sasaki, Evolution of the electronic structure of a ferromagnetic metal: Case of
SrRuO$_3$, Phys. Rev. B {\bf 80}, 035106 (2009).
\bibitem{scirep} S. Ryee and M. J. Han, Magnetic ground state of SrRuO$_3$ thin film and applicability of standard first-principles approximations to metallic magnetism, Sci. Rep. { \bf 7}, 4635 (2017).
\bibitem{zay} A. T. Zayak, X. Huang, J. B. Neaton, and K. M. Rabe, Structural, electronic, and magnetic properties of SrRuO$_ 3$ under epitaxial strain, Phys. Rev. B { \bf 74}, 094104 (2006).

\bibitem{sro}C. W. Jones,P. W. Battle, P Lightfoot, W.T.A. Harrison, The structure of SrRuO$_3$ by time-of-flight neutron powder diffraction. Acta Crystallogr, Sec. C: Cryst. Struct. Commun. {\bf 45}, 365-367 (1989).



\bibitem{dmft} L. Si, Z. Zhong, J. M. Tomczak, and K. Held, Route to room-temperature ferromagnetic ultrathin SrRuO$_3$ films, Phys. Rev. B { \bf 92}, 041108(R) (2015).

\bibitem{wan1} N. Marzari, A. A. Mostofi, J. R. Yates, I.  Souza, and D. Vanderbilt,  Maximally localized Wannier functions: Theory and applications, Rev. Mod. Phys. { \bf 84}, 1419-1475 (2012).
\bibitem{wan2}  A. A. Mostofi, \textit{et al.} An updated version of wannier90: A tool for obtaining maximally-localised Wannier functions, Comput. Phys. Commun. {\bf 185}, 2309-2310 (2014).
\bibitem{v3citation} J-P. Hanke, F. Freimuth, S. Blügel, and Y. Mokrousov, Higher-dimensional Wannier functions of multiparameter Hamiltonians, Phys. Rev. B {\bf 91}, 184413 (2015).
\bibitem{freimuth} F. Freimuth, Y. Mokrousov, D. Wortmann, S. Heinze, and S. Blügel, Maximally localized Wannier functions within the FLAPW formalism, Phys. Rev. B {\bf 78}, 035120 (2008).


\bibitem{lie} A. Liebsch, H. Ishida, Subband filling and mott transition in Ca$_{2-x}$Sr$_x$RuO$_4$, Phys. Rev. Lett. { \bf 98}, 216403 (2007).

\bibitem{khom} K. I. Kugel and D. I. Khomskii, Sov. The Jahn-Teller effect and magnetism: transition metal compounds, Phys. Usp. {  \bf 25}, 231 (1982).
\bibitem{Bruno} P. Bruno, Tight-binding approach to the orbital magnetic moment and magnetocrystalline anisotropy of transition-metal monolayers, Phys. Rev. B {\bf 39}, 865(R) (1989).

\bibitem{wang} X. Wang, J. R. Yates, I. Souza, and D. Vanderbilt, Ab initio calculation of the anomalous Hall conductivity by Wannier interpolation, Phys. Rev. B { \bf 74}, 195118 (2006).
\bibitem{yao} Y. Yao, \textit{et al.} First Principles Calculation of Anomalous Hall Conductivity in Ferromagnetic bcc Fe, Phys. Rev. Lett. { \bf 92}, 037204 (2004).
\bibitem{hcp-co}E.Roman, Y. Mokrousov, and I. Souza, Orientation Dependence of the Intrinsic Anomalous Hall Effect in hcp Cobalt, Phys. Rev. Lett. {\bf 103}, 097203 (2009).

\bibitem{Zhang} P. Zhang, A. Das,E. Barts, M. Azhar, L. Si, K. Held, M. Mostovoy, and T. Banerjee, Robust skyrmion-bubble textures in SrRuO$_3$ thin films stabilized by magnetic anisotropy, arXiv:2001.07039 (2020).

\bibitem{korea2} B. Sohn, \textit{et al.} Sign-tunable anomalous Hall effect induced by symmetry-protected nodal structures in ferromagnetic perovskite oxide thin films, arXiv:1912.04757 (2019).


\end{thebibliography}
\end{document}